\documentclass[aps,prl,reprint,superscriptaddress,showpacs,preprintnumbers]{revtex4-1}
\usepackage{amsmath}
\usepackage{natbib}
\usepackage{hyperref}
\usepackage{epsfig}

\begin{document}
\preprint{FREIBURG-PHENO-2011-008}
\preprint{Nikhef\,2011-014}
\preprint{MPP-2011-55}
\preprint{DESY 11-079}

\title{NLO QCD corrections to the production of two bottom-antibottom pairs
       at the LHC}

\author{Nicolas Greiner}
\affiliation{Department of Physics, %
	University of Illinois at Urbana-Champaign, %
	Urbana IL, 61801, USA}
\author{Alberto Guffanti}
\affiliation{Physikalisches Institut, %
	Albert-Ludwigs-Universit\"at, %
	79104~Freiburg, Germany}
\author{Thomas Reiter}
\affiliation{Nikhef, 1098~XG Amsterdam, The Netherlands}
\affiliation{Max-Planck-Institut f\"ur Physik, %
        80805~M\"unchen, Germany}
\author{J\"urgen Reuter}
\affiliation{DESY, %
        22607 Hamburg, %
	Germany}
\affiliation{Physikalisches Institut, %
	Albert-Ludwigs-Universit\"at, %
	79104~Freiburg, Germany}

\date{\today}

\begin{abstract}
We report the results of a computation of the full next-to-leading order QCD corrections to 
the production of two $b\bar{b}$ pairs at the LHC. 
This calculation at the parton level provides predictions for well separated $b$-jets. 
The results show that the next-to-leading order corrections lead to an enhancement of the 
cross-section for the central scale choice by roughly 50\% with respect to the leading order result. 
The theoretical uncertainty estimated by variation of the renormalization and factorization scales 
is strongly reduced by the inclusion of next-to-leading order corrections.
\end{abstract}


\maketitle

\section{Introduction}
The search for the Higgs boson, and more in general the study of the Electroweak Symmetry breaking 
mechanism, is a major goal of the experiments at the LHC collider at CERN. 
In various extensions of the Standard Model the signature of two light Higgs bosons decaying in 
two pairs of $b$-quarks, $hh\rightarrow b\bar{b}b\bar{b}$, is a viable channel for the Higgs Boson
discovery. Examples of these models are the Minimally Supersymmetric Standard Model (MSSM) for large 
values of $\tan\beta$ and moderate~$m_A$~\cite{RichterWas:1996ak,RichterWas:1997gi,Dai:1995cb,
Dai:1996rn}, hidden valley scenarios where the decay of hadrons of an additional gauge group can 
produce additional $b$-jets~\cite{Bern:2008ef,Strassler:2006im} and two Higgs doublet models. The 
possibility of measuring the Higgs self-coupling through $H\rightarrow hh\rightarrow b\bar{b}b\bar{b}$ 
has been investigated in~\cite{Lafaye:2000ec}. This and other related studies, however show that such 
a measurement would be extremely difficult, primarily due to the large Standard Model background. 
The precise knowledge of the $b\bar{b}b\bar{b}$ final state within the Standard Model is therefore an 
important factor for the success of these measurements.

Because of its importance this process has been added to the Les Houches wish list of relevant
next-to-leading order calculations~\cite{Bern:2008ef}.

In an earlier publication~\cite{Binoth:2009rv}, we presented the next-to-leading order (NLO) 
QCD corrections to the production of $b\bar{b}b\bar{b}$ via quark-antiquark annihilation. 
In the present Letter we complete the existing work including the gluon initiated contributions 
and present the results for the full NLO QCD corrections to $pp\rightarrow b\bar{b}b\bar{b}$ at 
the LHC.

We show that the inclusion of the NLO corrections reduces the unphysical scale 
dependence of the leading order (LO) prediction greatly, improving the precision of this prediction 
and allowing a better estimation of the Standard Model background to possible New Physics signals in
this channel.

\section{Method}

A complete NLO QCD description requires the calculation of the $2\to4$ subprocesses 
$q\bar{q}\to b\bar{b}b\bar{b}$ and $gg\to b\bar{b}b\bar{b}$ at the tree and the one-loop level as 
well as the $2\to5$ particle processes $q\bar{q}\to b\bar{b}b\bar{b}g$, $gg\to b\bar{b}b\bar{b}g$ 
and $\overset{{\scriptscriptstyle(}{\scriptstyle-}{\scriptscriptstyle)}}{q}%
g\to b\bar{b}b\bar{b}%
\overset{{\scriptscriptstyle(}{\scriptstyle-}{\scriptscriptstyle)}}{q}$ at tree~level.

We sum over four massless quark flavours $q\in\{u,d,s,c\}$ in the initial state. Neglecting the
contribution from initial state $b$-quarks is justified by the smallness of the $b$ parton distribution
function (PDF) with respect to the other quark PDFs. Moreover the fact that the gluon-gluon channel
is the dominant contribution at LHC energies further reduces the relative importance of the quark 
channels. We treat the $b$-quarks as massless, which is a very good approximation for LHC kinematics 
also due to the cuts imposed in order for the final state $b$-quarks to be detected and separated in 
phase space. Effects of the heavy top quark are neglected altogether in the final result after having
shown that they are numerically not~important.

The LO and the real radiation matrix elements are generated using MadGraph~\cite{Stelzer:1994ta}. 
For the subtraction of the infrared singularities we use Catani-Seymour dipoles~\cite{Catani:1996vz},
supplemented with a slicing parameter $\alpha$ as proposed in~\cite{Nagy:1998bb,Nagy:2003tz},
implemented in the MadDipole package~\cite{Frederix:2008hu,Frederix:2010cj}. 

As described in our earlier work~\cite{Binoth:2009rv}, we compute the one loop corrections to 
scattering matrix elements using an approach based on Feynman diagrams. 
The code for the numerical evaluation of the virtual corrections is generated using the automated 
one-loop matrix element generator \texttt{golem-2{.}0}~\cite{Binoth:2008gx,Reiter:2009kb,Reiter:2009dk}
which employs QGraf~\cite{Nogueira:1991ex}, Form~\cite{Vermaseren:2000nd}, the Form library 
Spinney~\cite{Cullen:2010jv} and the code generator Haggies~\cite{Reiter:2009ts} at intermediate 
levels of the diagram and code generation. The reduction and evaluation of the loop integrals is 
performed using the \textsc{Samurai}~\cite{Mastrolia:2010nb} and OneLoop~\cite{vanHameren:2010cp} 
packages~respectively.


The integration over phase space is carried out using MadEvent~\cite{Maltoni:2002qb} and it has
been split up in independent parts in order to optimize the computational time~required. 

The first contribution consists of the real emission matrix element supplemented with the subtraction 
terms. The integration of this contribution over the corresponding 13-dimensional phase space is one 
of the main computational bottlenecks of such a calculation. This integration has been performed 
using up to $2\cdot 10^9$ phase space points. For the $q \bar{q}$ and $gg$ subprocess the evaluation 
of a single phase space point requires the evaluation of 30 subtraction terms for each partonic channel
and an additional 10 subtraction terms are needed for the $qg$ channel. This means that a substantial 
fraction of CPU time is spent calculating the dipole contributions. In such a situation the use of a
value smaller than one for the slicing parameter $\alpha$, as proposed 
in~\cite{Nagy:1998bb,Nagy:2003tz}, speeds up substantially the computation by avoiding that each 
subtraction term is evaluated for each phase-space point. Besides the reduction of the computational 
time per point this setting has a second advantage. If not close to a singularity the integrand is 
given just by the real emission matrix element. Close to a singularity, where also subtraction terms 
are calculated, these subtraction terms per definition have the same kinematical structure as the real 
emission matrix element which is not necessarily true for an arbitrary point in phase space. 
So for each point the integrand is either exactly the real emission matrix element or something with 
the same structure but with one singularity subtracted. But as this is an integrand where our 
integration routine is optimized for, choosing a value for $\alpha$ smaller than one leads to an 
improvement of the convergence of the integral. In our calculation we set $\alpha=0.01$.

The second contribution to the integration combines the tree-level contribution and the integrated 
subtraction terms.
The virtual matrix element is integrated over phase space by reweighting a sample unweighted Born 
level events, as described in~\cite{Binoth:2008gx}. 
This leads to a considerable reduction of the required CPU-time since less phase space points have 
to be evaluated. For the results shown below, event samples consisting of $10^4-10^5$ unweighted
events have been used. The LO event samples used for the reweighting have been generated with 
MadEvent~\cite{Maltoni:2002qb} and WHIZARD ~\cite{Moretti:2001zz,Kilian:2007gr}.

In order to establish the correctness of the results obtained we have performed a number of 
non-trivial tests. The dipole contributions for single phase space points and at the phase space 
integration level have been compared with the HELAC code~\cite{Cafarella:2007pc,Czakon:2009ss} 
and agreement has been established up to double precision accuracy for single phase space points and
within integration errors for the integrated results. 
The phase space integration of the dipole contributions is validated by checking the independence 
of the result of the slicing parameter $\alpha$.
Also the cancellation of the single and double poles between the virtual amplitude and the integrated 
subtraction terms has been verified.
Finally, the virtual matrix element computation for a single phase space point has been 
compared to the result published in~\cite{vanHameren:2009dr}. In order to perform this comparison the 
contribution from top quark loops has been added, even though it is neglected in the results presented
in the following section. Our result is in agreement with the result of~\cite{vanHameren:2009dr}, 
providing a very strong test of our virtual contributions computations.

\section{Results}

In the following we consider the process $pp\to b\bar{b}b\bar{b}+X$ at the LHC at a center of mass 
energy of $\sqrt{s}=14\,\mathrm{TeV}$.
The final state jets are defined by applying the $k_T$-algorithm as explained in~\cite{Blazey:2000qt} 
with a radius in $R$-space of $0.8$. More precisely, the jet algorithm requires exactly four $b$-jets
in the final state for the event to be accepted.
All jets are required to lie within a rapidity range of $\vert\eta(b_j)\vert<2{.}5$ and to have a
transverse momentum~$p_T(b_j)>30\,\mathrm{GeV}$.
We impose a separation cut between the jets of 
$\Delta R(b_i,b_j)=\sqrt{(\phi_i-\phi_j)^2+(\eta_i-\eta_j)^2}>0{.}8$.
All results have been obtained using the CTEQ6M parton distribution functions~\cite{Pumplin:2002vw} 
with two-loop running of $\alpha_s$ both for the LO and the NLO cross-section evaluations 
and~$\alpha_s(M_Z)=0.118$.

The unphysical renormalization and factorization scales are usually chosen to be in the vicinity of 
the typical scale of the process. For processes where heavy particles such as top-quarks or 
$W/Z$-bosons are involved the masses of these particles provide a natural choice. 
In our case, dealing with massless particles only there is no such scale, the only scale involved in this
process is the $p_T$-cut imposed to define the $b$-jets. In this respect the process considered here 
is similar to the production of four light jets. 
In ~\cite{Nagy:2003tz} it has been shown that the average transverse momentum $p_T$ of a jet is a good 
choice for the production of three jets in hadron-hadron collisions. On the basis of the scale choice 
we made earlier for the quark initiated case~\cite{Binoth:2009rv}, we define the central scale to be
\begin{equation}
  \label{eq:centralscale}
  \mu_0=\frac{1}{4}\sqrt{\sum_ip_{T,i}^2}\;,
\end{equation}
which turns out to be of the same order of magnitude
as the average $p_T$ of the jets.

\begin{figure}[ht]
  \begin{center}
    \includegraphics[height=6cm]{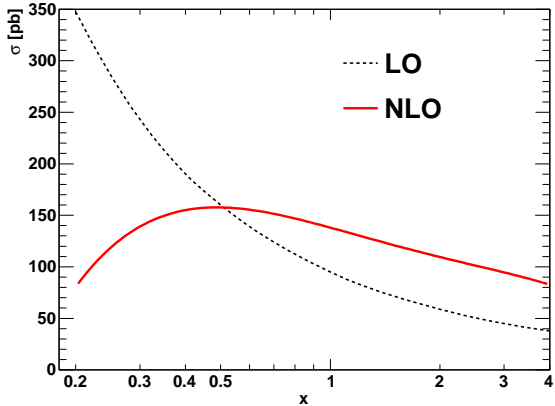}
  \end{center}
  \caption{Total cross section as a function of the scale $\mu=\mu_r=\mu_F=x\cdot \mu_0$.
    Renormalization and factorization scale are varied in the same direction.}
  \label{scale}
\end{figure}
In Figure \ref{scale} we plot the variation of the total cross section for the production of two 
bottom-antibottom pairs at the LHC, with the cuts described previously, when the renormalization 
scale $\mu_r$ and the factorization $\mu_F$ are varied together, with $x$ defined as the ratio to the
the central scale, $\mu_r=\mu_F= x \cdot \mu_0$. 

If we set the renormalization and factorization scale to the value~$\mu_0$
as in Eq.~\eqref{eq:centralscale} we find for
the total cross section with the cuts described above
\begin{equation}
  \sigma^{NLO}_{pp\to b\bar{b}b\bar{b}}= 140.48\; \pm \;0.64\; \text{pb}\;.
\end{equation}
This means that for our preferred choice of scales we find that the inclusion of the NLO contribution 
leads to an increase of nearly 50\% of the total cross section with respect to the LO result of
$\sigma^{LO}= 94.88\; \pm \; 0.14\; \text{pb}$.
However one observes that, as expected, the dependence of the result on the unphysical scales is 
strongly reduced in the NLO result with respect to the leading order one.

\begin{figure}[ht]
  \begin{center}
    \includegraphics[height=6cm]{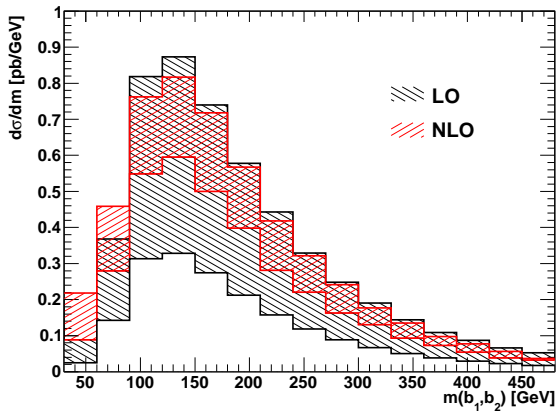}
  \end{center}
  \caption{Invariant mass distribution of the two $b$-jets with the
    highest~$p_T$.
    The black shaded area denotes the tree level contribution, the red
    area denotes the NLO cross-section. The error bands for both histograms
    are determined by a scale variation between $\mu_0/2$ and $2\mu_0$.
  }
  \label{mb1b2}
\end{figure}
In Figure \ref{mb1b2} we plot the invariant mass distribution of the two $b$-jets with the highest 
transverse momentum. The error bands are obtained by a variation of the scales between $\mu_0/2$ and 
$2\mu_0$. With respect to the LO result one observes a shift of the distribution to lower energies 
due to the inclusion of the radiative corrections.
\begin{figure}[ht]
  \begin{center}
    \includegraphics[height=6cm]{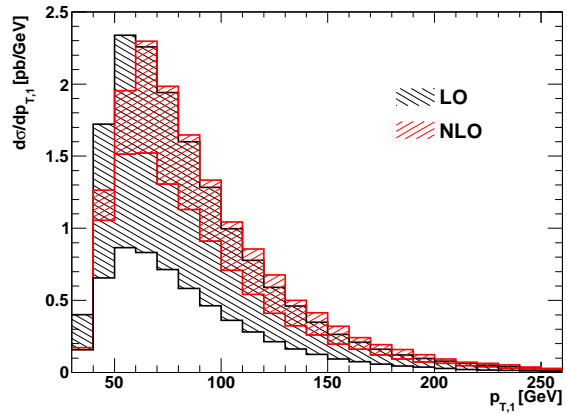}
  \end{center}
  \caption{$p_T$ distribution of the hardest jet. The error bands are defined
    as in Figure~\ref{mb1b2}.
  }
  \label{pt1}
\end{figure}
The $p_T$ distribution of the jet with the highest $p_T$ is shown in Figure~\ref{pt1}. 
Here, the radiative corrections enhance the distribution at higher momentum. 
Both Figure~\ref{mb1b2} and Figure~\ref{pt1} show a significant reduction of the error induced 
by scale uncertainties on differential distributions. Moreover, the distortion in the shapes of 
differential distributions when going from LO to NLO suggests that the application of a global 
$K$-factor is not sufficient in order to accurately describe the higher-order effects.

A complete phenomenological study of the production of two bottom-antibottom pairs at the 
LHC, including the study of PDF uncertainties and the effects of varying the cuts on the final
state $b$-jets is beyond the scope of the present Letter and will be the subject of an upcoming
publication.

\section{Conclusions}

We have calculated the next-to-leading order QCD corrections to the production of two 
bottom-antibottom quark pairs at the LHC.
This calculation has been implemented in a highly automated framework for the computation of
NLO QCD corrections (the {\tt golem-2.0} framework) which is based on a Feynman diagrammatic 
approach for the evaluation of virtual corrections implemented in the \textsc{Samurai} and OneLoop 
packages, interfaced to the Madgraph/Madevent and MadDipole programs for the evaluation of the 
leading-order and dipole subtraction contributions and the phase space integration.

The inclusion of the NLO corrections leads to a significant reduction of the uncertainties due to
unphysical scale dependence of the LO result, while enhancing the cross section by 50\% for our
central scale choice. Furthermore, we have shown that the radiative corrections lead to changes in 
the overall shape of the distributions, which cannot be accounted for in a reliable way by a simple 
rescaling of the leading order predictions.

This reduced theoretical uncertainty improves the prospects for the use of the $b\bar{b}b\bar{b}$ 
channel in searches of Higgs bosons in various extensions of the the standard model like SUSY, two
Higgs doublet models or hidden valley models. 

\begin{acknowledgments}
{\bf Acknowledgments}: The authors would like to thank Gudrun Heinrich for useful discussion
and for her support in cross-checking parts of the amplitude.
N.G. wants to thank Francesco Tramontano for helpful advice. N.G and A.G would like to thank Nikhef 
for kind hospitality.
The work of T.R. was supported by the Dutch Foundation for Fundamental Research
on Matter~(FOM), project FORM 07PR2556
and by the Alexander von Humboldt Foundation,
in the framework of the Sofja Kovaleskaja Award Project
"Advanced Mathematical Methods for Particle Physics",
endowed by the German Federal Ministry of Education and Research.
N.G was supported by the U.~S.~Department of Energy
under contract No.~DE-FG02-91ER40677.

This paper is dedicated to the memory of our friend Thomas Binoth who initiated this project and
has always been a driving force in our collaboration.

\end{acknowledgments}

\bibliography{article_4b}

\begin{thebibliography}{10}

\bibitem{RichterWas:1996ak}
Elzbieta Richter-Was.
\newblock Minimal supersymmetric standard model higgs rates and backgrounds in
  atlas.
\newblock {\em Int. J. Mod. Phys. A}, 13(CERN-TH-96-111):1371--1494. 174 p, Apr
  1996.

\bibitem{RichterWas:1997gi}
Elzbieta Richter-Was and Daniel Froidevaux.
\newblock {MSSM Higgs searches in multi-b jet final states at the LHC}.
\newblock {\em Z.Phys.}, C76:665--676, 1997.

\bibitem{Dai:1995cb}
J.~Dai, J.F. Gunion, and R.~Vega.
\newblock {Detection of the minimal supersymmetric model Higgs boson $H^0$ in
  its $h^0 h^0\to 4b$ and $A^0A^0\to 4b$ decay channels}.
\newblock {\em Phys.Lett.}, B371:71--77, 1996.

\bibitem{Dai:1996rn}
J.~Dai, J.F. Gunion, and R.~Vega.
\newblock {Detection of neutral MSSM Higgs bosons in four b final states at the
  Tevatron and the LHC: An update}.
\newblock {\em Phys.Lett.}, B387:801--803, 1996.

\bibitem{Bern:2008ef}
Z.~Bern et~al.
\newblock {The NLO multileg working group: Summary report}.
\newblock 2008.

\bibitem{Strassler:2006im}
Matthew~J. Strassler and Kathryn~M. Zurek.
\newblock {Echoes of a hidden valley at hadron colliders}.
\newblock {\em Phys. Lett.}, B651:374--379, 2007.

\bibitem{Lafaye:2000ec}
R.~Lafaye, 2~Miller, D.J., M.~Muhlleitner, and S.~Moretti.
\newblock {Double Higgs production at TeV colliders in the minimal
  supersymmetric standard model}.
\newblock 2000.

\bibitem{Binoth:2009rv}
T.~Binoth, N.~Greiner, A.~Guffanti, J.~Reuter, J.-Ph. Guillet, et~al.
\newblock {Next-to-leading order QCD corrections to pp to b anti-b b anti-b + X
  at the LHC: the quark induced case}.
\newblock {\em Phys.Lett.}, B685:293--296, 2010.

\bibitem{Stelzer:1994ta}
T.~Stelzer and W.F. Long.
\newblock {Automatic generation of tree level helicity amplitudes}.
\newblock {\em Comput.Phys.Commun.}, 81:357--371, 1994.

\bibitem{Catani:1996vz}
S.~Catani and M.H. Seymour.
\newblock {A General algorithm for calculating jet cross-sections in NLO QCD}.
\newblock {\em Nucl.Phys.}, B485:291--419, 1997.

\bibitem{Nagy:1998bb}
Zoltan Nagy and Zoltan Trocsanyi.
\newblock {Next-to-leading order calculation of four-jet observables in
  electron positron annihilation}.
\newblock {\em Phys. Rev.}, D59:014020, 1999.

\bibitem{Nagy:2003tz}
Zoltan Nagy.
\newblock {Next-to-leading order calculation of three jet observables in hadron
  hadron collision}.
\newblock {\em Phys. Rev.}, D68:094002, 2003.

\bibitem{Frederix:2008hu}
Rikkert Frederix, Thomas Gehrmann, and Nicolas Greiner.
\newblock {Automation of the Dipole Subtraction Method in MadGraph/MadEvent}.
\newblock {\em JHEP}, 0809:122, 2008.

\bibitem{Frederix:2010cj}
R.~Frederix, T.~Gehrmann, and N.~Greiner.
\newblock {Integrated dipoles with MadDipole in the MadGraph framework}.
\newblock {\em JHEP}, 1006:086, 2010.

\bibitem{Binoth:2008gx}
T.~Binoth, A.~Guffanti, J.-Ph. Guillet, G.~Heinrich, S.~Karg, et~al.
\newblock {Precise predictions for LHC using a GOLEM}.
\newblock {\em Nucl.Phys.Proc.Suppl.}, 183:91--96, 2008.

\bibitem{Reiter:2009kb}
Thomas Reiter.
\newblock {Automated Evaluation of One-Loop Six-Point Processes for the LHC}.
\newblock 2009.

\bibitem{Reiter:2009dk}
Thomas Reiter.
\newblock {An Automated Approach for q anti-q to b anti-b b anti-b at
  Next-to-Leading Order QCD}.
\newblock 2009.

\bibitem{Nogueira:1991ex}
P.~Nogueira.
\newblock Automatic feynman graph generation.
\newblock {\em J.Comput.Phys.}, 105:279--289, 1993.

\bibitem{Vermaseren:2000nd}
J.A.M. Vermaseren.
\newblock {New features of FORM}.
\newblock 2000.

\bibitem{Cullen:2010jv}
Gavin Cullen, Maciej Koch-Janusz, and Thomas Reiter.
\newblock {Spinney: A Form Library for Helicity Spinors}.
\newblock 2010.

\bibitem{Reiter:2009ts}
Thomas Reiter.
\newblock {Optimising Code Generation with haggies}.
\newblock {\em Comput.Phys.Commun.}, 181:1301--1331, 2010.

\bibitem{Mastrolia:2010nb}
P.~Mastrolia, G.~Ossola, T.~Reiter, and F.~Tramontano.
\newblock {Scattering AMplitudes from Unitarity-based Reduction Algorithm at
  the Integrand-level}.
\newblock {\em JHEP}, 1008:080, 2010.

\bibitem{vanHameren:2010cp}
A.~van Hameren.
\newblock {OneLOop: For the evaluation of one-loop scalar functions}.
\newblock 2010.

\bibitem{Maltoni:2002qb}
Fabio Maltoni and Tim Stelzer.
\newblock {MadEvent: Automatic event generation with MadGraph}.
\newblock {\em JHEP}, 0302:027, 2003.

\bibitem{Moretti:2001zz}
Mauro Moretti, Thorsten Ohl, and Jurgen Reuter.
\newblock {O'Mega: An optimizing matrix element generator}.
\newblock 2001.

\bibitem{Kilian:2007gr}
Wolfgang Kilian, Thorsten Ohl, and Jurgen Reuter.
\newblock {WHIZARD: Simulating Multi-Particle Processes at LHC and ILC}.
\newblock 2007.

\bibitem{Cafarella:2007pc}
Alessandro Cafarella, Costas~G. Papadopoulos, and Malgorzata Worek.
\newblock {Helac-Phegas: a generator for all parton level processes}.
\newblock {\em Comput. Phys. Commun.}, 180:1941--1955, 2009.

\bibitem{Czakon:2009ss}
M.~Czakon, C.G. Papadopoulos, and M.~Worek.
\newblock {Polarizing the Dipoles}.
\newblock {\em JHEP}, 0908:085, 2009.

\bibitem{vanHameren:2009dr}
A.~van Hameren, C.~G. Papadopoulos, and R.~Pittau.
\newblock {Automated one-loop calculations: a proof of concept}.
\newblock {\em JHEP}, 09:106, 2009.

\bibitem{Blazey:2000qt}
Gerald~C. Blazey, Jay~R. Dittmann, Stephen~D. Ellis, V.Daniel Elvira, K.~Frame,
  et~al.
\newblock {Run II jet physics}.
\newblock pages 47--77, 2000.

\bibitem{Pumplin:2002vw}
J.~Pumplin, D.R. Stump, J.~Huston, H.L. Lai, Pavel~M. Nadolsky, et~al.
\newblock {New generation of parton distributions with uncertainties from
  global QCD analysis}.
\newblock {\em JHEP}, 0207:012, 2002.

\end{thebibliography}
\end{document}